\documentclass[12pt,preprint]{aastex}

\begin{document}

\title{High Density Molecular Gas in the IR-bright Galaxy System VV~114}
\author {Daisuke Iono\altaffilmark{1,2}, Paul T. P. Ho\altaffilmark{1,3}, Min S. Yun\altaffilmark{2},  Satoki Matsushita\altaffilmark{3}, Alison B. Peck\altaffilmark{1}, Kazushi Sakamoto\altaffilmark{1}}
\altaffiltext{1}{Harvard-Smithsonian Center for Astrophysics, 60 Garden Street, Cambridge, MA 02138}
\altaffiltext{2}{Department of Astronomy, University of Massachusetts, Amherst, MA 01002}
\altaffiltext{3}{Academia Sinica Institute of Astronomy and Astrophysics, P.O. Box 23-141, Taipei 106, Taiwan, R.O.C.}
\begin{abstract}

New high resolution CO~(3--2) interferometric map of the IR-bright interacting galaxy system VV~114 observed with the Submillimeter Array (SMA) reveal a substantial amount of warm and dense gas in the IR-bright but optically obscured galaxy, VV~114E, and the overlap region connecting the two nuclei.  A $1.8 \times 1.4$ kpc concentration of CO~(3--2) emitting gas with a total mass of $4\times 10^9 M_\odot$ coincides with the peaks of NIR, MIR, and radio continuum emission found previously by others, identifying the dense fuel for the AGN and/or the starburst activity there.  Extensive CO~(2--1) emission is also detected, revealing detailed distribution and kinematics that are consistent with the earlier CO~(1--0) results.  The widely distributed molecular gas traced in CO~(2--1) and the distributed discrete peaks of CO~(3--2) emission suggest that a spatially extended intense starbursts may contribute significantly to its large IR luminosity.  These new observations further support the notion that VV~114 is approaching its final stage of merger, when violent central inflow of gas triggers intense starburst activity possibly boosting the IR luminosity above the ultraluminous threshold. 

\end{abstract}

\keywords{galaxies: interactions --- galaxies: kinematics and dynamics --- galaxies: individual (VV~114)}

\section{Introduction}

Observations of luminous and ultraluminous infrared galaxies (LIRGs/ULIRGs) are the key to understanding the formation and evolution of galaxies and their associated star forming environment \citep{sanders96,genzel00}.    Tidal interaction between two or more gas-rich progenitor galaxies is largely responsible for radially transporting the gas to the central kpc, condensing the gas and triggering subsequent starburst activity there \citep{mihos96}.  It is now widely accepted that the elevated level of infrared luminosity originates from the reprocessed emission from the dust particles surrounding the starburst or AGN.  Based on a number of similarities, the infrared bright galaxies are thought to be the local analogs of the high redshift sub-mm sources discovered using SCUBA on JCMT during the last decade \citep{blain02}, and understanding the nearby LIRGs/ULIRGs population is an important step toward better understanding the sub-mm galaxy phenomenon.   

Investigation of LIRGs/ULIRGs using molecular gas at the low J-transitions (CO~(1--0) at 2.6 mm and CO~(2--1) at 1.3 mm) were carried out extensively in the past \citep{bryant99,downes98}.  While these CO transitions were believed to be a good tracer of the optically thick dense gas identifying the extent of the starburst region and the distribution of the fuel for such activity, more recent studies suggest that the diffuse inter-clump medium may dominate the CO luminosity in these low J-transition CO emission \citep{downes98}. Since these two J-transitions require only a small difference in excitation conditions, the large scale distribution of the molecular gas inferred is usually quite similar \citep{downes98}.  The $J=3\rightarrow 2$ transition of CO has a higher excitation temperature (33 K) and critical density ($\sim 10^4$ cm$^{-3}$), making it a better tracer of the warmer and denser molecular gas of the starburst regions.  In this letter, we present  high resolution CO~(3--2) and CO~(2--1) interferometric maps of the IR-bright galaxy system VV~114 observed with the Submillimeter Array (SMA)\footnote{The Submillimeter Array is a joint project between the Smithsonian Astrophysical Observatory and the Academia Sinica Institute of Astronomy and Astrophysics, and is funded by the Smithsonian Institution and the Academia Sinica} \citep{ho04} to trace the distribution of the warmer dense gas and its relation to the colder, more diffuse molecular gas.  

VV~114 is a gas rich \citep[M$_{\rm H_2} = 5.1 \times 10^{10} \rm M_{\odot}$; ][YSK94 hereafter]{yun94} nearby ($\rm D = 77$ Mpc) interacting system with high infrared luminosity \citep[$\rm L_{\rm IR} = 4.0 \times 10^{11} \rm L_{\odot}$; ][]{soifer87}.  The projected nuclear separation between the two optical galaxies (VV~114E and VV~114W) is $\sim 6$ kpc.  \citet{frayer99} found a large amount of dust ($\rm M_{dust} = 1.2 \times 10^8 \rm M_{\odot}$) distributed across the galaxy with dust temperature of $20 - 25$ K.  About half of the warmer dust traced in the MIR is associated with VV~114E, where both compact (nuclear region) and extended emission are found \citep{lefloch02}.  The MIR spectrum also shows a sign of an AGN in VV~114E \citep{lefloch02}.  \citet{herrero02} detected abundant \ion{H}{2} regions in VV~114E and in the overlap region using the narrow-band Pa$\alpha$ images.  \citet{scoville00} imaged the near infrared emission using NICMOS on-board HST and found that the highly optically obscured VV~114E \citep{knop94} is the brighter of the two in the near infrared.  Far-UV imaging using STIS found several hundred young star clusters in VV~114W, while no UV emission was found in VV~114E \citep{goldader02} which suggests that most of the activity in VV~114E is obscured by dust and not visible in the UV emission.  

\begin{deluxetable}{lccc}
\tablecaption{Observed and Derived Properties of VV~114\label{tbl-1}}
\tablewidth{0pt}
\tablehead{
\colhead{Parameter} & \colhead{CO(3-2)} & \colhead{CO(2-1)} & \colhead{CO(1-0)\tablenotemark{a}}  
} 
\startdata
$\theta$ (FWHM)\tablenotemark{b}\\
~~arcsecond  & $2.9 \times 2.2$ & $5.0 \times 3.0$  &  $6.5 \times 3.7$  \\
~~kpc & $1.1 \times 0.8$ & $1.9 \times 1.2$ &  $2.4 \times 2.2$ \\
Cont. RMS (mJy)\\
~~Observed & 50 & 5 & -\\
~~Theoretical & 20 & 5 & -\\
Peak (J2000)\\
~~R.A.  & $01^{\rm h} 07^{\rm m} 47.5^{\rm s}$ & $01^{\rm h} 07^{\rm m} 47.2^{\rm s}$ &  $01^{\rm h} 07^{\rm m} 47.3^{\rm s}$ \\
~~Decl.  & $-17^{\circ} 30' 25.0''$ & $-17^{\circ} 30' 25.0''$ & $-17^{\circ} 30' 25.7''$ \\
$S \Delta v$ (Jy km/s)\tablenotemark{c} & 2275 & 1620 & 674\\
$v_{sys}$ (km/s)&  6020 & 6027 & 6040\\ 
$\Delta v_{FWZI}$ (km/s) & 330 & 520 & 520\\ 
\enddata 
\tablenotetext{a}{from \citet{yun94}}
\tablenotetext{b}{Natural weighting}
\tablenotetext{c}{The integrated flux is subject  to a $20\%$ (CO~(2--1)) uncertainty and factor of $\sim 2$ overestimate (CO~(3--2)), mainly because of the uncertainties in flux and gain calibration.}
\end{deluxetable}

\section{Observations \& Results}

VV~114 was observed under good weather conditions ($\tau_{230} \sim 0.08$) 
in CO~(3--2) and CO~(2--1) using six 6-meter diameter elements of 
the SMA on September 30, 2003 and on September 12, 2003, respectively.  
The receivers were tuned to 339.0 GHz (LSB) and 226.0 GHz (LSB) which 
resulted in a primary beam size of $35''$ and $52''$ (FWHM) for CO~(3--2) 
and CO~(2--1).  The array was in the ``compact'' configuration 
which includes the shortest and longest projected baseline of 
10 and 65 meters, respectively.  The largest detectable structure 
are thus $18''$ and $27''$ in the CO~(3--2) and CO~(2--1) lines, 
and the missing short spacing information may have
substantially impacted the continuum and CO (3--2) maps (see below). 
These observations were made during the testing and commissioning phase of the construction of the array, and extra calibration steps were taken to scrutinize the array performance.  
Two nearby quasars, 0132$-$169 and 0050$-$094, were observed every 25 minutes to track the instrumental gain, and absolute flux calibration was performed by observing Uranus.  The telescope pointing drifted significantly (10-20$''$) during the track, and the resulting amplitude error primarily limits the overall dynamic range of the images to $\lesssim 20$.
Initial data calibration was carried out using the Caltech
millimeter array software package MIR
which is modified for SMA. The calibrated data were imaged using NRAO software package AIPS \citep{moo96}.  The spectrometer was configured with a total bandwidth of 960 MHz with a 0.8 MHz channel spacing.   
The final spectral channel maps were made by smoothing the data 
to a velocity resolution of about 22 km/s.  Finally, the astrometry was examined by calibrating 0050$-$094 (secondary) with 0132$-$169 (primary), and we estimate a positional uncertainty of $\sim 0.3''$.  A summary of the observations is given in Table~\ref{tbl-1}.  Further technical descriptions of the SMA and its calibration schemes are found in \citet{ho04}.

Continuum images are constructed by averaging the entire 960 MHz bandwidth
of the image sideband data, and neither the 1.3 mm nor the $850 \micron$ continuum is detected with $3\sigma$ upper limits of 15 mJy and 150 mJy, respectively (see Table~\ref{tbl-1}).   
The JCMT/SCUBA submm continuum images by \citet{frayer99} suggest a source extent of $\sim 30''$ (11 kpc) in diameter.  From their dust model fit to the $450 \micron$ and $850 \micron$ measurements, we estimate a total 1.3 mm flux density of $\sim100$ mJy.  Smoothing the SMA data to the angular resolution of the JCMT does not reveal any emission feature, and a significant fraction of the dust emission may arise from a smooth structure larger than our shortest baseline ($\theta\gtrsim27''$).  Assuming the extent of the dust emission is similar to that of the CO~(1--0) emission, we derive a $3\sigma$ upper limit in 1 mm continuum of 120 mJy. 

\begin{figure*}[t]
\epsscale{1}
\plotone{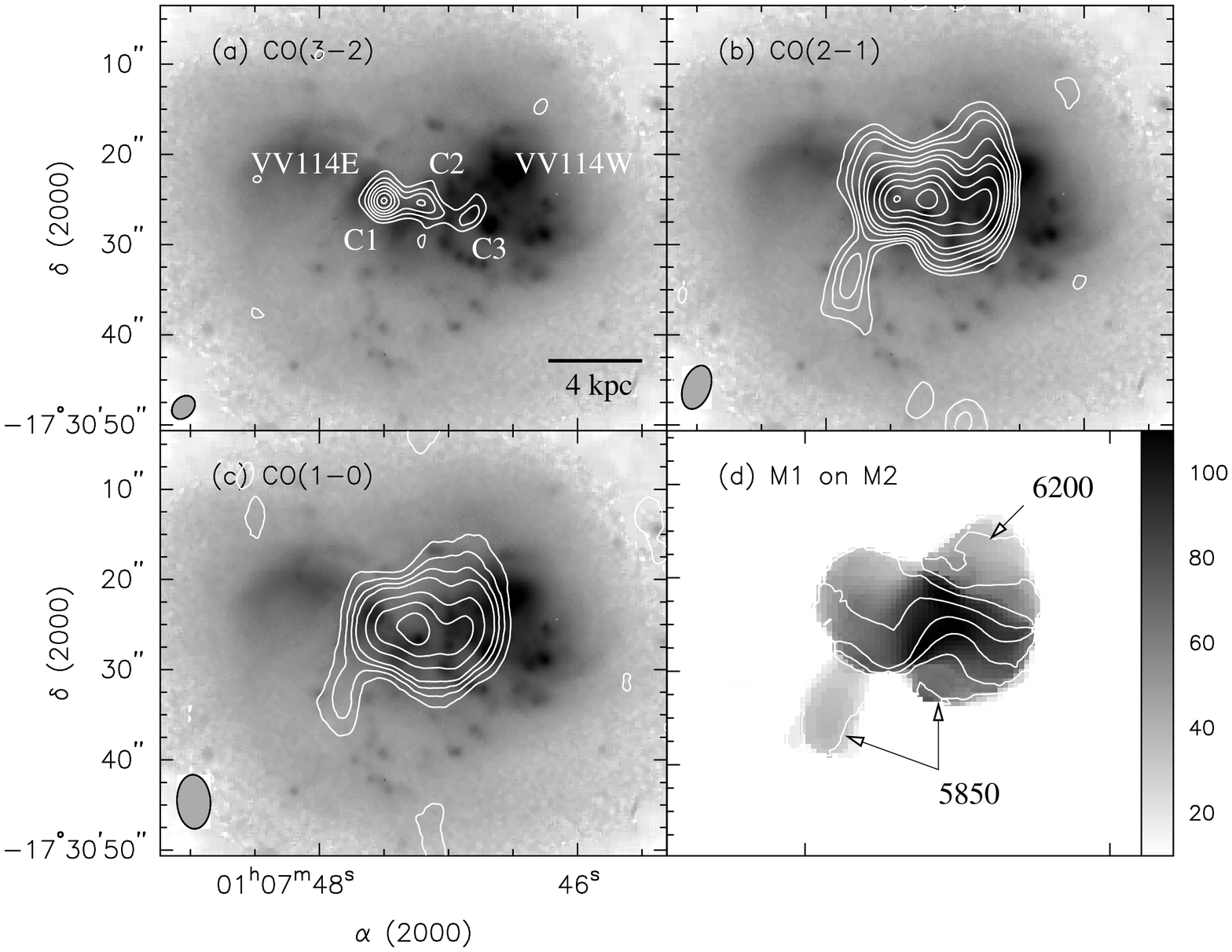}
\caption{(a) CO~(3--2), (b) CO~(2--1), and (c) CO~(1--0) (YSK94) maps overlaid over R-band optical image.  The contour levels represent 4,6,8,10,12,14,16 $\times \sigma$ (40 Jy km/s) for CO~(3--2) ($\theta = 2.9'' \times 2.2''$), 6,8,12,16,20,30,40,50,60 $\times \sigma$ (5.0 Jy km/s) for CO~(2--1) ($\theta = 5.0'' \times 3.0''$), and 4,6,8,10,15,20,25,30 $\times \sigma$ (3.5 Jy km/s) for CO~(1--0) ($\theta = 6.5'' \times 3.7''$).  The three resolved CO~(3--2) molecular complexes are labeled C1, C2 and C3 from east to west, where C1 is the most dominant of the three ($\sim 50\%$ of the total flux).  Each of the CO images are centered at the phase center of each observations.  No primary beam correction is applied since the source size is smaller than the primary beam even at 330 GHz. (d)  The CO~(2--1) velocity distribution contours (M1) (units in km/s) overlaid over the velocity dispersion (M2) map in gray scale which ranges from 10 to 110 km/s.  The velocity contours are equally spaced in 50 km/s intervals.  This figure was made using the WIP software package \citep{morgan95}.
\label{fig1}}
\end{figure*}

\subsection{CO(3--2) Emission}

Our new SMA CO~(3--2) image, shown in Figure~\ref{fig1}a, reveals a   
$5.0 \times 2.0$ kpc bar-like morphology with two peaks (C1 and C3) occupying the two ends and a third peak C2 located near the center of the bar.  A similar structure is seen in both lower J transitions (see Fig.~\ref{fig1} and compare the higher contours of CO~(2--1) with the lower contours of CO~(3--2)).   
More than 50\% of the total CO~(3--2) flux is detected in the compact molecular concentration labeled C1 in the IR-bright dusty eastern galaxy (VV114E), coincident with the NIR \citep{doyon95, scoville00}, MIR \citep{lefloch02} and radio continuum emission \citep{condon90}.  Assuming optically thick emission and the standard galactic $CO-H_2$ conversion \citep{sanders91}, its inferred molecular gas mass is $4\times 10^9 M_\odot$, and this accounts for nearly 40\% of its virial mass ($10^{10}$ M$_\odot$ for a deconvoled source size of $1.4 \times 1.8$ kpc and $\Delta v = 100$ km/s), similar to what is commonly seen in the nuclear regions of other infrared luminous galaxies \citep{scoville97,downes98}.  The peak C2 roughly coincides with the brightest peaks in CO~(2--1) and CO~(1--0), and the peak C3 extends to the part of the region of VV~114W where copious clusters of young stars have been observed previously \citep{herrero02, goldader02}.  Despite the strong CO~(3--2) emission seen in VV~114E, the main disk of the optically bright western galaxy (VV~114W) shows little emission, possibly resolved out by the interferometer.  The nuclear region of VV~114E exhibits higher line intensity ratios ($r_{21} = I_{21}/I_{10}$ and $r_{31} = I_{32}/I_{10}$) than the rest of the system, suggesting the presence of significantly warmer, denser gas.  The inference of high gas temperature is further supported by the observed peak brightness temperature of $\Delta T \sim10$ K averaged over the 1 kpc beam area (i.e., $T\sim100$ K for a filling factor $f\sim0.1$).  

The absence of baselines shorter than 10 meters and the resulting limited sensitivity to structures $\gtrsim 18''$ in extent may account for much of the extended structure missing in the CO (3--2) image. For spectral line observations, this spatial filtering is not as severe as one naively calculates since coherent bulk motions such as galactic rotation substantially reduces the effective angular sizes of the emitting regions in individual velocity channels.  Multi-isotopic, multi-transition analysis of molecular tracers have indicated a multi-phase medium with a significant contribution by sub-thermally excited diffuse gas in the turbulent nuclear starburst regions and thus a lower conversion factor between the CO luminosity and the total molecular gas mass \citep[see ][]{aalto95,downes98}. Therefore, some of the observed morphological differences from the lower J transitions (which are more easily excited) may reflect real differences in the physical properties of gas.  A more definitive characterization of the distribution and physical properties of the warm, dense gas traced in CO (3--2) will require obtaining a single dish measurement in the future.  The total observed line flux of 2275 Jy km/s translates to a total molecular gas mass of $2\times 10^{10} M_\odot$, which is about 40\% of the total gas mass inferred from the CO (1--0) measurement by (YSK94), and this demonstrates clearly both the large gas mass and the highly concentrated nature of the dense ($n\ge 10^4$ cm$^{-3}$) molecular gas in VV~114.

\begin{figure}[t]
\epsscale{1}
\plotone{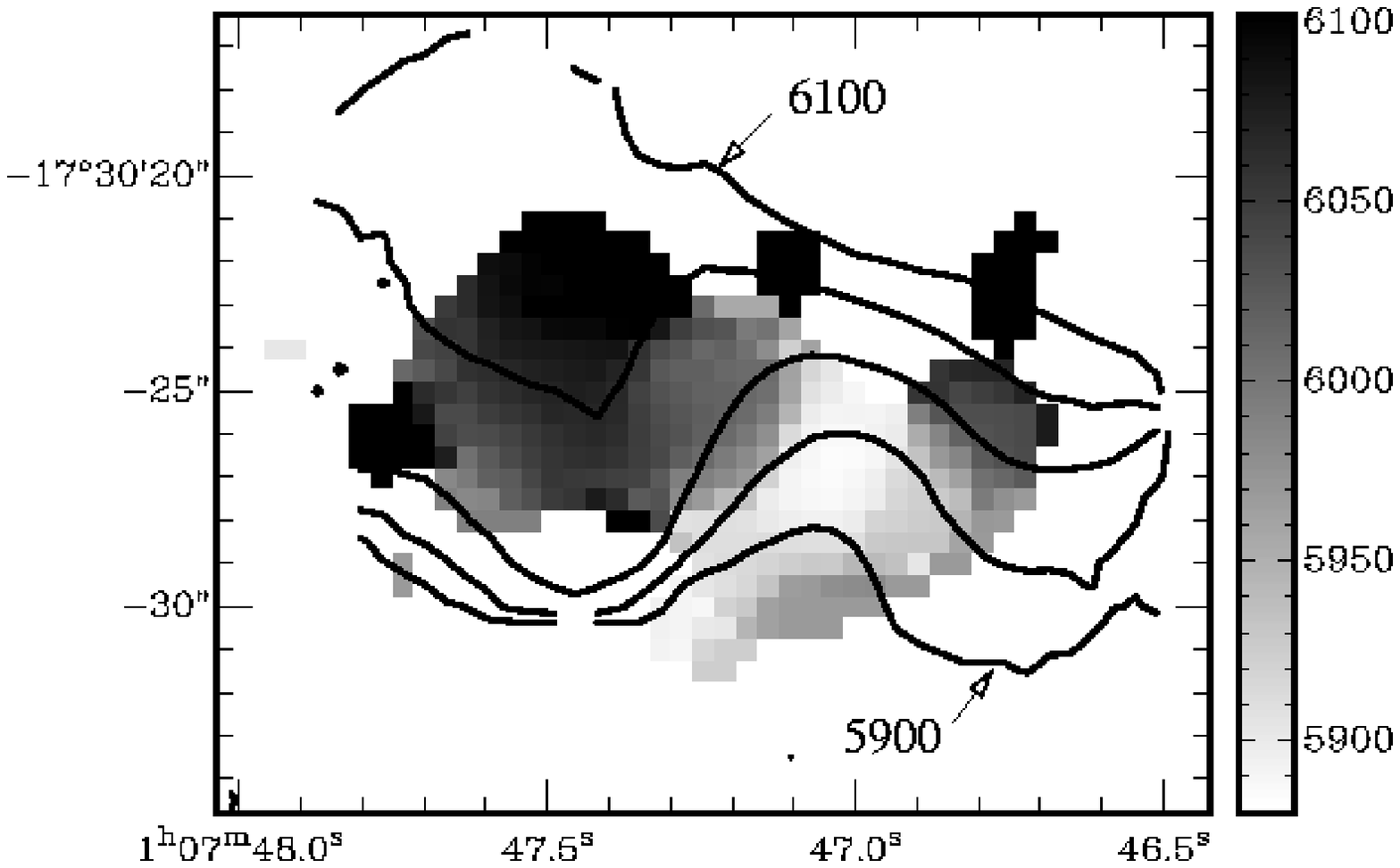}
\caption{The CO~(3--2) velocity distribution is shown in gray scale (in km/s) overlaid with the velocity contours of the CO~(2--1) emission.  The contours range from 5900 to 6100 km/s in steps of 50 km/s.
\label{fig2}}
\end{figure}

The clumpy spatial distribution of the high density gas makes the CO~(3--2) kinematics appear complex (see Figure~\ref{fig2}).  A rotation-like velocity gradient is seen from the southwest-northeast with a noticeably steeper gradient near the C2.  The direction of the overall velocity gradient is consistent with that in CO~(1--0) (YSK94) and in CO~(2--1) (Fig.~\ref{fig1}d).  A large amount of extended emission may be missing by our observations, and the derived velocity field should be interpreted with a caution.  Both the excitation requirement for the CO (3--2) emission and the spatial filtering by the interferometer may be isolating the kinematics of the densest gas, possibly in the deepest parts of the evolving potential.  The total linewidth at zero intensity is 330 km/s, much narrower than those traced in CO (1--0) and CO (2--1), and the linewidth in VV~114E alone spans over 100 km/s.  The peak C1 is most likely associated with VV~114E, and we thus infer the systemic velocity of 6052 km/s for VV~114E.

\subsection{CO(2--1) Emission}

The velocity integrated CO~(2--1) map is shown in Figure~\ref{fig1}b.  The CO~(2--1) emission extends widely across the two disks comprising VV~114, showing a high degree of  similarity to the CO~(1--0) (YSK94; also Fig.~\ref{fig1}c), including both the bar-like structure $5 \times 3$ kpc in extent dominating the central region and a long (3 kpc) molecular tail extending out from the southeastern edge with no obvious optical counterpart.  The higher angular resolution of the CO~(2--1) map makes subtle features appear more enhanced, but the overall distribution is very similar to the CO~(1--0) map.  The excellent correspondence of the two is not surprising since both transitions are easily excited by the physical conditions typical of cold, dense ($T=10-20$ K, $n \sim 10^3$ cm$^{-3}$) gas.  The global line intensity ratio $r_{21} = I_{21}/I_{10} = 2.4\pm0.7$ is much smaller than the expected value of 4 for fully thermalized, optically thick emission from a single phase gas.  This is a further indication that sub-thermal excitation is an important concern for CO transitions in this object.

The gas kinematics traced in CO(2--1) and CO(1--0) lines are also nearly identical.  The total linewidth of the CO~(2--1) emission is 520 km/s which is comparable to the linewidth of the CO~(1--0) emission.  By taking the median velocity of the emission channels, the CO~(2--1) systemic velocity of VV~114 as a whole is estimated at 6027 km/s.  A steep velocity gradient is seen in both CO~(2--1) and CO~(1--0) where the southeastern molecular tail connects to the main component of molecular gas in VV~114, but a smaller gradient is seen along the 3 kpc length farther south. This may suggest that the molecular tail is tidally driven, and its kinematics are dominated by the angular momentum of the part of the galaxy from which it was initially perturbed (see below).
The large velocity dispersion is seen near C2 in both CO~(2--1) and CO~(1--0) emission, possibly marking the new dynamical center of the merger (YSK94).  Alternatively, the large velocity dispersion may imply a tidally induced radial motion seen along the line of sight.  

\section{Discussion}

While several hundred young star clusters have been identified in VV~114W \citep{goldader02}, most of the current vigorous activity is seen in the highly obscured galaxy, VV~114E, as identified by its infrared, radio and sub-mm continuum emission.  The origin of the intense IR emission, starburst or AGN or both, in VV~114 is still subject to debate.  While \citet{lefloch02} suggests that 40\% of the MIR emission is associated with the compact nuclear AGN in VV~114E, different diagnostics \citep[i.e. NIR spectroscopy, ][]{doyon95} suggest that the IR emission is mostly dominated by young stars.  The three massive, compact gas complexes identified by our CO (3--2) observations mark the locations of the densest concentrations of gas and the sites of current activities fueled by this gas. The dominant peak C1 in VV~114E appears to mark the location of the most intense current activity.  The distributed nature of CO~(3--2) emission and the presence of several massive gas concentrations provide the clearest evidence yet that distributed intense starburst activity may provide a significant fraction of the large IR luminosity in VV~114. 

Our new CO (3--2) and CO (2--1) observations strengthen the earlier suggestion that VV~114 is a late stage merger ready to undergo more massive bursts of star formation in the near future.  The widely extended CO~(2--1) and CO~(1--0) emission across the two galaxies implies that the cold and dense gas have begun to decouple from the gravitational potential of the host galaxies, and such decoupling only occurs in the well advanced stages ($t \sim 5 \times 10^8$ years since the initial collision) before the coalescence results in a massive elliptical galaxy \citep[e.g., ][]{mihos96}.  The three discrete peaks traced in CO (3--2) emission identify the massive and compact molecular gas concentrations condensing out of the inflowing gas. 
By comparing the observed distribution of gas and optical morphology to those predicted in simulations \citep{mihos96,iono04}, we estimate that the final coalescence may occur within the next $\sim 10^8$ years.  
The CO~(1--0) and CO~(2--1) gas distribution of VV~114 is very similar to what is seen in the IR-bright galaxy NGC~6090 \citep{bryant99, wang03}.  There the widely extended gas is seen over the edge-on and face-on system separated by 3.4 kpc, where not only the CO~(1--0) and CO~(2--1) but also the CO~(3--2) gas peaks in the overlap region \citep{wang03}.  Thus, both VV~114 and NGC~6090 appears to be undergoing a similar transition where the bulk of the gas is in the process of decoupling from the host galaxies and funneling toward the dynamical center of the new combined potential, possibly resulting in the ultraluminous ($L_{IR}\ge 10^{12} L_\odot$) phase, following the scenario advocated by \citet{yun94}.

The authors would like to thank J. Hibbard for kindly supplying the R-band image of VV~114.  D. I. and M.Y. are grateful for the warm and extensive support of the SMA group at Hilo during D.I.'s pre-doc residency at the SMA site, and the countless stimulating discussions with the members of the SMA.  This and other science observations were made possible only with the significant contributions from the staff members, engineers, operators and administrators at Hilo, Cambridge and ASIAA in Taiwan.   This research is partly supported by the Faculty Research Grant at the University of Massachusetts and the National Science Foundation grant AST 97-25951.


\begin{thebibliography}{fun}
\bibitem[Aalto et al.(1995)]{aalto95} Aalto, S., Booth, R. S., Black, J. H. \& Johansson, L. E. B., 1995 A\&A, 300, 369
\bibitem[Alonso-Herrero, Rieke \& Rieke(2002)]{herrero02} Alonso-Herrero, A., Rieke, G. H., \& Rieke, M. J. 2002, ApJ, 124, 166
\bibitem[Blain et al.(2002)]{blain02} Blain, A. W., Smail, I., Ivison, R. J., Kneib, J. P., \& Frayer, D. T.,  2002, PhR, 369, 111 
\bibitem[Bryant \& Scoville(1999)]{bryant99} Bryant, P. M., \& Scoville, N. Z. 1999, AJ, 117, 2632 
\bibitem[Condon et al.(1990)]{condon90} Condon, J. J., Helou, G., Sanders, D. B., \& Soifer, B. T. 1990, ApJS, 73, 359
\bibitem[Downes \& Solomon(1998)]{downes98} Downes, D, \& Solomon, P. M. 1998, ApJ, 364, 615
\bibitem[Doyon et al.(1995)]{doyon95} Doyon, R., Nadeau, D., Joseph, R. D.,  Goldader, J. D., Sanders, D. B., \& Rowlands, N. 1995, ApJ, 450, 111
\bibitem[Frayer et al.(1999)]{frayer99} Frayer, D. T., Ivison, R. J., Smail, I., Yun, M. S., \& Armus, L. 1999, ApJ, 118, 139
\bibitem[Genzel \& Cesarsky(2000)]{genzel00} Genzel, R., \& Cesarsky, C. J. 2000, ARAA, 38, 761
\bibitem[Goldader et al.(2002)]{goldader02} Goldader, J. D., Gerhardt, M., Heckman, T. M., Seiber, M., Sanders, D. B., Calzetti, D., \& Steidel, C. C. 2002, ApJ, 568, 651
\bibitem[Ho, Moran \& Lo(2004)]{ho04} Ho, P. T. P, Moran, J. M. \& Lo, K. Y. 2004 ApJL, this issue
\bibitem[Iono, Yun \& Mihos(2004)]{iono04} Iono, D., Yun, M. S., \& Mihos, C. J. in preparation
\bibitem[Knop et al.(1994)]{knop94} Knop, R. A., Soifer, B. T., Graham, J. R., Matthews, K., Sanders, D. B., \& Scoville, N. Z. 1994, ApJ, 107, 920
\bibitem[Le Floc'h et al.(2002)]{lefloch02} Le Floc'h, E., Charmandaris, V., Laurent, O., Mirabel, I. F., Gallais, P., Sauvage, M., Vigroux, L., \& Cesarsky, C. 2002, A\&A, 391, 417 
\bibitem[Mihos \& Hernquist(1996)]{mihos96} Mihos, C. J., \& Hernquist, L. 1996, ApJ, 464, 641 
\bibitem[Morgan (1995)]{morgan95} J. A. Morgan. WIP - An Interactive Graphics Software Package, in: Astronomical Data Analysis Software and Systems IV, ed. R. A. Shaw, H. E. Payne, and J. J. E. Hayes. PASP Conf Series 77, 129 (1995)
\bibitem[Sanders \& Mirabel(1996)]{sanders96}Sanders, D. B., \& Mirabel, I. F. 1996, ARAA, 34, 749
\bibitem[Sanders, Scoville \& Soifer(1991)]{sanders91}Sanders, D. B., Scoville, N. Z. \& Soifer, B. T. 1991, 370, 158
\bibitem[Scoville, Yun, \& Bryant (1997)]{scoville97} Scoville, N. Z., Yun, M. S., \& Bryant, P. M. 1997, ApJ, 484, 702
\bibitem[Scoville et al.(2000)]{scoville00} Scoville et al. 2000, AJ, 119, 991
\bibitem[Soifer et al.(1987)]{soifer87} Soifer, B. T., Sanders, D. B., Madore, B. F., Neugebauer, G., Danielson, G. E., Elias, J. H., Lonsdale, C. J. \& Rice, W. L. 1987, ApJ, 320, 238
\bibitem[van Moorsel, Kemball \& Greisen(1996)]{moo96} van Moorsel, G., Kemball, A., \& Greisen, E. 1996, in ASP Conf Ser. 101, Astronomical Data Analysis Software and Systems V, ed. G. H. Jacoby \& J. Barnes (San Francisco: ASP), 37
\bibitem[Yun, Scoville \& Knop(1994)]{yun94} Yun, M. S., Scoville, N. Z., \& Knop, R. A. 1994, ApJL, 430, 109 
\bibitem[Wang et al.(2004)]{wang03} Wang, J. Z. et al. 2004, ApJL, this issue
\end{thebibliography}
\end{document}